\documentclass[draftcls,12pt,onecolumn,oneside]{IEEEtran}
\usepackage{pifont}
\usepackage{times,amsmath,color,amssymb,graphicx,epsfig,cite,psfrag,subfigure}
\usepackage{mathrsfs}
\usepackage{amsfonts}

\ifCLASSOPTIONcompsoc
\else
\fi
\ifCLASSINFOpdf
\else
\fi
\begin{document}
\title{Incentive Mechanism Design for Heterogeneous Peer-to-Peer Networks: A Stackelberg Game Approach}
%
%
%
%

\author{Xin~Kang,~\IEEEmembership{Member,~IEEE,}
        Yongdong~Wu,~\IEEEmembership{Member,~IEEE}

\IEEEcompsocitemizethanks{\IEEEcompsocthanksitem Part of the results
of this paper have been presented in IEEE ICC 2013, Budapest,
Hungary \cite{XKangICC_2013}.}
\IEEEcompsocitemizethanks{\IEEEcompsocthanksitem X. Kang and Y. Wu
are with Institute for Infocomm Research, 1 Fusionopolis Way,
$\sharp$21-01 Connexis, South Tower, Singapore 138632. \protect
Email: \{xkang, wydong\}@i2r.a-star.edu.sg.}
\thanks{}}

%
%

\markboth{IEEE Transactions on Mobile Computing}%
{Shell \MakeLowercase{\textit{et al.}}: Bare Demo of IEEEtran.cls
for Computer Society Journals}

\IEEEcompsoctitleabstractindextext{%
\begin{abstract}
With high scalability, high video streaming quality, and low
bandwidth requirement, peer-to-peer (P2P) systems have become a
popular way to exchange files and deliver multimedia content over
the internet. However, current P2P systems are suffering from
``free-riding" due to the peers' selfish nature. In this paper, we
propose a credit-based incentive mechanism to encourage peers to
cooperate with each other in a heterogeneous network consisting of
wired and wireless peers. The proposed mechanism can provide
differentiated service to peers with different credits through
biased resource allocation. A Stackelberg game is formulated to
obtain the optimal pricing and purchasing strategies, which can
jointly maximize the revenue of the uploader and the utilities of
the downloaders. In particular, peers' heterogeneity and selfish
nature are taken into consideration when designing the utility
functions for the Stackelberg game. It is shown that the proposed
resource allocation scheme is effective in providing service
differentiation for peers and stimulating them to make contribution
to the P2P streaming system.
\end{abstract}

}

\maketitle

\IEEEdisplaynotcompsoctitleabstractindextext

%
\IEEEpeerreviewmaketitle

\section{Introduction}
With the rapid development of peer-to-peer (P2P) communication
technologies, P2P networks have become a popular way to exchange
files and deliver multimedia content over the internet due to their
low bandwidth requirement, good video streaming quality, and high
flexibility. However, current P2P systems greatly rely on
volitionary resource contribution from individual peers and do not
enforce any compulsory contribution from these peers. This directly
leads to the well-known ``free-riding" problem, which refers to the
phenomenon that a peer consumes free service provided by other peers
without contributing any its resources to the P2P network. This
tremendously degrades the performance of P2P systems, especially P2P
multimedia streaming systems which have high requirements on time
delay and data rate. Free-riding is common in P2P networks due to
peers' selfish nature and the limited network resources. Most peers
only want to maximize their own benefits without caring about the
overall performance of the whole P2P community. It is reported in
\cite{EAdar_200010} that more than $70\%$ P2P users do not share any
file in Gnutella system. Therefore, to enhance the performance of
P2P networks, effective incentive mechanisms need to be put in place
to stimulate the cooperation between peers and encourage them to
make contribution to the P2P system.

On the other hand, recent advances in wireless communications
technologies (3G/4G networks) and smart phones have enabled the
development of mobile version of P2P applications for smart phones,
such as PPtv \cite{PPTV} and PPStream \cite{PPStream}. People use
these mobile P2P applications to watch movies, watch dramas, or
listen to music when traveling on buses and metros. Due to the
convenience, mobile P2P users are increasing dramatically nowadays.
As compared to the wired P2P users, mobile P2P users are more
selfish due to the high cost of mobile data. Thus, there is also a
compelling need to design effective incentive mechanisms for mobile
P2P applications.

The existing incentive mechanisms for P2P systems are mainly
designed to work in wired networks. For the heterogeneous networks
with both wired and wireless nodes, these incentive mechanisms may
not work well due to the differences between the wired nodes and the
wireless nodes. For example, the computing capability of the
wireless nodes (such as smart phones and tablet PCs) is usually
weaker than that of the wired nodes (such as desktop PCs, and
workstations). Thus, incentive mechanisms with high complexity may
not be suitable for mobile applications. It is true that there exist
high-end smartphones with high-end four-core or eight-core
processors. However, incentive mechanisms with high complexity are
still not preferred on these mobile devices since the high
complexity computing can drain out the devices' batteries fast. In
addition, the connection bandwidth of the wireless nodes is usually
less than that of the wired nodes. This should be taken into
consideration when designing the incentive mechanism to achieve
relative fairness. However, to the best of our knowledge, most of
the existing work fails to do this. All these differences between
the wireless and wired nodes pose new challenges to the design of
the incentive mechanism for the heterogeneous networks.

In this paper, we propose a credit-based incentive mechanism for
heterogeneous networks with both wired and wireless nodes. We
consider a P2P streaming network where each peer can serve as an
uploader and a downloader at the same time. When a peer uploads data
chunks to other peers, it can earn certain credits for providing the
service. When a peer downloads data chunks from other peers, it has
to pay certain credits for consuming the resource. A peer's net
contribution to the network is reflected by its accumulated credits.
A Stackelberg game is formulated to provide differentiated service
to peers with different credits. Particularly, peers' heterogeneity
and selfish nature are taken into consideration when designing the
utility functions.

The main contributions and key results of this paper are summarized
as follows.
\begin{itemize}
  \item A credit-based incentive mechanism based on Stackelberg
games is proposed for P2P streaming networks. To the best of our
knowledge, this is the first work that applies the Stackelberg game
to the incentive mechanism design for P2P streaming networks.
 \item Peers' heterogeneity is taken into consideration
  when designing the utility functions for the Stackelberg game. Thus, our incentive mechanism can be applied to
 heterogeneous P2P networks with wired and wireless peers having different connection bandwidths.
  \item The selfish nature of peers is taken into consideration
  when designing the utility functions for the Stackelberg game, i.e.,
  every peer is a strategic player with the aim to maximize its
  own benefit. This makes our incentive mechanism perform well
  in a P2P network environment with non-altruistic peers.
  \item The optimal pricing strategies for the uploader and the
  optimal purchasing strategies for the downloader are both derived.
  The Stackelberg equilibrium is then obtained and shown to be unique and Pareto-optimal.
  \item Two fully distributed implementation schemes are proposed based on the
  obtained theoretical results. It is shown that each of these
  schemes has its own advantages.
  \item The impact of peer churn on the proposed incentive mechanism
  is analyzed. It is shown that the proposed mechanism can adapt to
  dynamic events such as peers joining or leaving the network.
\end{itemize}

The remaining parts of this paper are organized as follows. In
Sections \ref{sec-relatedworks} and \ref{Sec-Framework}, we present
the related work and describe our system model. In Sections
\ref{Sec-SE formulation} and \ref{Sec-Op solution}, we present the
problem formulation and its optimal solution. In Sections
\ref{Sec-implementation} and \ref{Sec-DynamicsAnalysis}, we propose
two implementation schemes and study the impact of peer churn on the
proposed schemes. Numerical results are given in Section
\ref{NumericalResults} to evaluate the performance of the proposed
schemes. Then, we discuss some possible extensions of this work in
Section \ref{Sec-futurework}. Section \ref{conclusions} concludes
the paper.

\section{Related Work}\label{sec-relatedworks}
A simple incentive mechanism for P2P systems is the ``tit-for-tat''
strategy, where peers receive only as much as they contribute. A
free rider that does not upload data chunks to other peers cannot
get data chunks from them and suffers from poor streaming quality.
Due to its simplicity and fairness, this scheme has been adopted by
BitTorrent \cite{BCohen_2003}. Though this strategy can increase the
cooperation between peers to a certain level, it is shown in
literature  \cite{SJun_2005, MPiatek_2007, MingluLi} that it may
perform poorly in today's internet environment due to the asymmetry
of the upload and download bandwidths.

Unlike the ``tit-for-tat'' strategy, which enforces compulsory
contribution from peers, another category of incentive mechanisms
stimulate peers to contribute to the system by indirect reciprocity
\cite{PGolle_200110,zhong2003sprite,YChu_200408,MFeldman_2004,yu2005incentive,
AHabib_200606,WSLin_201007,feldman2005overcoming,XSu_201009}. In
these incentive mechanisms, the contribution of each peer is
converted to a score which is then used to determine the reputation
or rank of the peer among all the peers in the network. Peers with a
high reputation are given  a certain priority in utilizing the
network resources, such as selecting peers or desirable media data
chunks. Therefore, peers with a high reputation have more
flexibility in choosing desired data suppliers and thus are more
likely to receive high-quality streaming. On the other hand, peers
with a low reputation have quite limited options in parent-selection
and thus receive low-quality streaming. Through this way, the P2P
systems can provide differentiated service to peers with different
reputation values. Hence, peers are motivated to contribute more to
the P2P system to earn a higher reputation.

Recently, game theory \cite{GameTheory1993} is found to be a
powerful tool to study strategic interactions among rational peers
and design incentive mechanisms to stimulate the cooperation among
peers for P2P streaming systems. This is due to the fact that peers
are selfish and strategic players in P2P streaming systems. It is
their inherent nature to maximize their payoffs while simultaneously
reducing their cost, i.e., enjoying a high quality streaming service
while consuming least of their own resource. Game theory has been
widely used in studying strategic interactions among these peers
\cite{mahajan2004experiences, GuptaR_2005,
CCourcoubetis_2006,PAntoniadis_2007,FuXie_2008,wang2003play,
sun2004slic, LinYao_2011,WSLin_200904,GTan_200807,RMa_200610}. In
\cite{mahajan2004experiences, GuptaR_2005,
CCourcoubetis_2006,PAntoniadis_2007,FuXie_2008}, the authors
discussed how to apply game theory to the design of incentive
mechanisms for P2P networks at a high level. It is pointed out that
straightforward use of results from traditional game theory do not
fit well with the requirements of P2P networks. The utility
functions must be customized for P2P networks. In
\cite{wang2003play}, a repeated static game called Cournot Oligopoly
game was formulated to model the interactions between peers, and an
incentive mechanism was proposed by analyzing  and solving  the
game. In \cite{sun2004slic}, a simple, selfish, link-based incentive
mechanism for unstructured P2P file sharing systems was proposed. It
was shown that a greedy approach is sufficient for the system to
evolve into a ``good'' state under the studied game model. In
\cite{LinYao_2011}, an incentive mechanism was proposed for P2P
networks based on the Bayes game. In \cite{WSLin_200904}, an
infinitely repeated game was formulated to analyze the interactions
between peers, and a so-called \emph{credit line} mechanism was
proposed to stimulate cooperation between peers. In
\cite{GTan_200807}, based on the \emph{first-price auction}
procedure, a payment-based incentive mechanism was proposed for P2P
streaming networks. Whereas, in \cite{RMa_200610}, a non-cooperative
competition game was used to provide service-differentiated resource
allocation between competing peers in a P2P network.

Different from these work, to the best of our knowledge, our work is
the first work that models the peers' interactions as a Stackelberg
game. Particularly, we take the peers' heterogeneity (wired/wireless
peers with different connection bandwidths) into consideration when
designing the utility functions for the Stackelberg game. Besides,
two distributed implementations of the mechanism with different
complexity are proposed to handle the difference in computing
capability between wired and wireless nodes.

\section{System Model}\label{Sec-Framework}
\begin{figure}[t]
        \centering
        \includegraphics*[width=7cm]{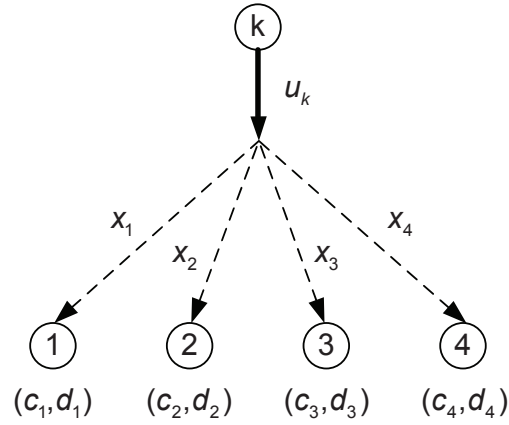}
        \caption{System Model}
        \label{Fig-System Model}
\end{figure}
In this paper, we consider a P2P streaming network where all the
peers can serve as the uploader and the downloader at the same time.
To eliminate the free-riding phenomenon and encourage cooperation
between peers, we introduce the concept of credit into the system,
where peers earn credits for providing service and
consume credits for receiving service. 
We assume that all the peers are selfish and rational. Their aim is
to maximize the credits that they can earn by fully utilizing their
available network resource. Each peer has the right to set up a
price for the service that it provides based on its own benefits.
For fairness considerations, we assume that the uploader can only
adopt the uniform pricing strategy, i.e., it cannot set different
prices for different peers for the same amount of bandwidth
allocation.

In this paper, the credit of peer $i$ is denoted by $c_i$. The
connection type (i.e., the download capacity) of peer $i$ is denoted
by $d_i$. The download bandwidth allocation for peer $i$ is denoted
by $x_i$. The upload bandwidth of peer $k$ is denoted by $u_k$. We
denote the set of peers that request data chunks from peer $k$ as
$S_k$.   To avoid trivial bandwidth allocation schemes, we assume
that $\sum_{i\in S_k} d_i>u_k$. As illustrated in Fig.
\ref{Fig-System Model}, the downloaders send their credits and
connection types to the the uploader together with their data
request. The uploader then decides the bandwidth price and allocates
the bandwidth to requesters based on their credits and connection
types. For example, suppose there are $100$ peers requesting data
chunks from peer $i$, but peer $i$ can only provide service to $20$
peers at the same time due to its limited upload bandwidth. Then,
peer $i$ can set up a high price that only $20$ peers can accept,
and the remaining $80$ peers will give up due to the high cost.
Through this way, peers with more credits are actually given a
higher priority in utilizing  network resources, and service
differentiation for peers with different credits is thus realized.
In this paper, the above service differentiation scheme is realized
by a Stackelberg resource allocation game which is investigated in
the following section.

\section{Stackelberg Game Formulation} \label{Sec-SE formulation}
A Stackelberg game is a strategic game that consists of a leader and
several followers competing with each other on certain resources.
The leader moves first and the followers move subsequently. In this
paper, we formulate the uploader that has media data chunks as the
leader, and the downloaders that request for media data chunks as
the followers. The uploader (leader) imposes a price on each unit of
bandwidth providing to each downloader. Then, the downloaders
(followers) determine their optimal download bandwidths to maximize
their individual utilities based on the assigned bandwidth price.
The Stackelberg Game consists of two parts: the uploading game at
the uploader side and the downloading game at the downloader side,
which are introduced in the following two subsections, respectively.

\subsection{Uploading Game Design}
Under the Stackelberg game model, for the uploader $k$, if we denote
its price on each unit of bandwidth providing to each downloader as
$\mu$, then its revenue maximization problem can be formulated as

 \underline{Uploading Game:}
\begin{align}
\max_{\mu> 0}~&\sum_{i\in S_k} \mu x_i,\label{Eq-SGame-Obj}\\
\mbox{s.t.}~~& \sum_{i\in S_k} x_i \le u_k,
\end{align}
where $x_i$ is the bandwidth that peer $i$ intends to purchase, and
$x_i$ is a function of the bandwidth price $\mu$, i.e.,
$x_i\triangleq f_i(\mu)$. $S_k$ denotes the set of peers that
request data chunks from peer $k$, and $u_k$ is the total available
upload bandwidth of peer $k$.

Under the Stackelberg game formulation, the amount of bandwidth that
peer $i$ intends to purchase is decreasing in the bandwidth price
$\mu$. 
On the other hand, it is observed from \eqref{Eq-SGame-Obj} that the
revenue of the uploader is the sum of products of the bandwidth
price and individual peer's purchased bandwidth. Therefore, the
uploader must carefully design its bandwidth pricing strategy in
order to maximize its revenue.

\subsection{Downloading Game Design}
At the downloader side, for each peer $i$ that requests data chunks
from the uploader, the utility maximization problem can be
formulated as

\underline{Downloading Game:}
\begin{align}
\max_{x_i\ge 0}~& c_i s_i-\mu x_i,\label{Eq-DownloadGame_obj}\\
\mbox{s.t.}~~&x_i \le d_i,
\end{align}
where $s_i\triangleq \log_2\left(1+\frac{x_i}{d_i}\right)$ is the
\emph{performance satisfaction factor} for peer $i$, $c_i$ is the
credits that peer $i$ has, and $d_i$ is the maximum download
bandwidth of peer $i$.

The performance satisfaction factor $s_i$ reflects the degree of
satisfaction or the ``happines'' of the downloader under the
received bandwidth $x_i$. A log function is adopted to model this
factor due to the fact that log functions are shown in literature to
be suitable to representing a large class of elastic data traffics
including the media streaming service
\cite{SShenker_199509,LingjieDuan2012}. When the received bandwidth
$x_i=0$, the satisfaction factor $s_i$ is equal to $0$, which
indicates that peer $i$ is unsatisfied with its system performance.
On the other hand, when the received bandwidth $x_i=d_i$, the
satisfaction factor $s_i$ is equal to $1$, which indicates that peer
$i$ is fully satisfied with its system performance. The degree of
satisfaction increases with the increase of the received bandwidth
$x_i$.

It is also observed from \eqref{Eq-DownloadGame_obj} that the
utility function of the downloader consist of two parts: $c_i s_i$
and $\mu x_i$.  $c_i s_i$ is the credits that peer $i$ is willing to
pay for the service it received, while $\mu x_i$ is the cost that
peer $i$ has to pay for obtaining the bandwidth $x_i$. Obviously,
with a larger bandwidth $x_i$, peer $i$ can obtain more satisfactory
system performance, and thus is willing to pay more credits. On the
other hand, the cost increases with the increase of the bandwidth
$x_i$. Therefore, optimal strategies are needed for a rational peer
to balance its cost and the achieved system performance in order to
maximize its utility.

\subsection{Stackelberg Equilibrium}
The uploading game and the downloading game together form a
Stackelberg game. The objective of this game is to find the
Stackelberg Equilibrium (SE) point(s) from which neither the leader
nor the followers have incentives to deviate. For the proposed
Stackelberg game, the SE is defined as follows.

\underline{\textbf{Definition 3.1:}} Let $\mu^*$ be a solution for
the uploading problem and $x_i^{*}$ be a solution for the
downloading game of the $i$th peer. Then, the point
$\left(\mu^*,\boldsymbol{x}^*\right)$ is a SE for the proposed
Stackelberg game if for any $\left(\mu,\boldsymbol{x}\right)$ with
$\mu>0$ and $\boldsymbol{x}\succeq \boldsymbol{0}$, the following
conditions are satisfied:
\begin{align}
U^{up}\left(\mu^*,\boldsymbol{x}^*\right)&\ge
U^{up}\left(\mu,\boldsymbol{x}^*\right),\\U^{down}_i\left(x_i^*,\mu^*\right)&\ge
U^{down}_i\left(x_i,\mu^*\right), \forall i,
\end{align}
where $U^{up}(\cdot)$ and $U^{down}(\cdot)$ are the utilities of the
uploading game and the downloading game, respectively.

For the proposed game in this paper, the SE can be obtained as
follows: For a given price $\mu$, the downloading game is solved
first. Then, with the obtained best response functions
$\boldsymbol{x}^*$ of the downloaders, we solve the uploading game
for the optimal price $\mu^*$.

\section{Optimal Resource Allocation Strategies} \label{Sec-Op solution}
In this section, we investigate the optimal resource allocation
strategies for the proposed Stackelberg game, i.e., the optimal
bandwidth allocation for the downloading game and the optimal
pricing strategy for the uploading game.
\subsection{Optimal Download Bandwidth}

For a given $\mu$, the optimal bandwidth $x_i^*$ for peer $i$ is
given in the following theorem.

\underline{\textbf{Theorem 4.1:}} For a given $\mu$, the optimal
solution for the downloading game is
\begin{align}\label{Eq-xi}
x_i^*=\left\{\begin{array}{cl}
             d_i, & \mbox{if}~\mu \le \frac{c_i}{2 d_i \ln2 },\\
             \frac{c_i}{\mu\ln2}-d_i, & \mbox{if} ~\frac{c_i}{2 d_i\ln2 }<\mu\le \frac{c_i}{d_i \ln2 },\\
             0, & \mbox{if}~\mu > \frac{c_i}{d_i\ln2 }.
           \end{array}
\right.
\end{align}
\begin{proof}
The Lagrangian of the downloading game can be written as
\begin{align}
L(x_i,\alpha, \beta)=c_i \log_2\left(1+\frac{x_i}{d_i}\right)-\mu
x_i-\alpha\left(x_i-d_i\right)+\beta x_i. \nonumber
\end{align}
where $\alpha$ and $\beta$ are the nonnegative dual variable
associated with the constraints.

The dual function is $q(\alpha, \beta)=\max_{x_i} L(x_i,\alpha,
\beta)$. The Lagrange dual problem is then given by $\min_{\alpha
\ge0, \beta \ge0} q(\alpha, \beta)$. The duality gap is zero for the
convex problem addressed here, and thus solving its dual problem is
equivalent to solving the original problem. Thus, the optimal
solutions needs to satisfy the following Karush-Kuhn-Tucker (KKT)
conditions \cite{kangJSAC_2011}:
\begin{eqnarray}
d_i\ge x_i^* \ge 0, \alpha\ge0, \beta \ge0,\\
\alpha\left(x_i^*-d_i\right)=0,\label{Eq-16}\\
\beta x_i^*=0,\label{Eq-17}\\
\frac{\partial L(x_i^*,\alpha, \beta)}{\partial x_i^*}
=\frac{c_i}{\left(d_i+x_i^*\right)\ln2}-\mu-\alpha+\beta=0.
\label{Eq-18}
\end{eqnarray}
From \eqref{Eq-18}, it follows
\begin{align}\label{Eq-19}
x_i^*=\frac{c_i}{\left(\mu+\alpha-\beta\right)\ln2}-d_i.
\end{align}

Suppose $x_i^*<d_i$ when $\mu\le\frac{c_i}{2d_i\ln2}$. Then, from
\eqref{Eq-16}, it follows $\alpha=0$. Therefore, \eqref{Eq-19}
reduces to $x_i^*=\frac{c_i}{\ln2\left(\mu-\beta\right)}-d_i$. Then
$ x_i^*<d_i$ results in $\frac{c_i}{2d_i\ln2}<\mu-\beta$. Since
$\beta \ge0$, it follows $\mu>\frac{c_i}{2d_i\ln2}$. This
contradicts the presumption. Therefore, from \eqref{Eq-16}, it
follows
\begin{align}x_i^*=d_i,~ \mbox{if}~ \mu\le\frac{c_i}{2d_i\ln2}.\end{align}

Similarly, we can prove that $x_i^*=0, ~\mbox{if}~\mu >
\frac{c_i}{d_i\ln2 }$, and $x_i^*=\frac{c_i}{\mu\ln2}-d_i,~\mbox{if}
~\frac{c_i}{2 d_i\ln2 }<\mu\le \frac{c_i}{d_i \ln2 }$.

Theorem 4.1 is thus proved.
\end{proof}

\underline{\textbf{Remark:}} It is observed from \eqref{Eq-xi} that
$x_i^*$ is a piecewise function of the price $\mu$. If the price
$\mu$ is very high, the optimal download bandwidth $x_i^*$ for peer
$i$ is $0$; if the price $\mu$ is very low, peer $i$ will download
at its maximum bandwidth. In general, $x_i^*$ is a decreasing
function of $\mu$. This indicates that the uploader can easily
control the bandwidth allocated to peer $i$ by controlling the price
$\mu$ assigned for peer $i$. Besides, some key observations obtained
from \eqref{Eq-xi} are listed as follows.

\begin{itemize}
  \item Under the same prescribed price $\mu$, comparing with the same
type (i.e., the same $d_i$) of peers with higher contributions
(i.e., more credits), low contributors are more likely to be
rejected from downloading. The uploader can easily reject a low
contributor $i$ from downloading by setting a price larger than
$\frac{c_i}{d_i\ln2 }$.
  \item Under the same prescribed price $\mu$, more bandwidth is
allocated to the peer with higher contributions for the same type of
peer. High contributors are more likely to download at their maximum
download bandwidth.
\end{itemize}

\subsection{Optimal Pricing Strategy}\label{Sec-OPS}
With the results obtained in \eqref{Eq-xi}, we are now ready for
solving the uploading game. Using $f_i(\mu)$ to denote the $x_i$
obtained in \eqref{Eq-xi}, the uploading game can be rewritten as
\begin{align}
\mbox{P1:}~~\max_{\mu> 0}~&\sum_{i\in S_k} \mu f_i(\mu),\\
\mbox{s.t.}~~& \sum_{i\in S_k} f_i(\mu) \le u_k. \label{P1-con}
\end{align}

The above problem is difficult to solve since $f_i(\mu)$ is a
piece-wise function of $\mu$. Therefore, to solve P1, we first
consider the two-peer scenario, and then extend the results to the
multi-peer scenario.

\subsubsection{Two-peer Scenario}
In this scenario, we consider the case that only two peers request
data chunks from the uploader, and we assume that they are sorted in
the order $\frac{c_1}{d_1}>\frac{c_2}{d_2}$. Then, the thresholds
given in \eqref{Eq-xi} of the two peers may have the following two
possible orders.

Case I: $\frac{c_1}{
d_1\ln2}>\frac{c_2}{d_2\ln2}>\frac{c_1}{2d_1\ln2}>\frac{c_2}{2d_2\ln2}$.

Case II:
$\frac{c_1}{d_1\ln2}>\frac{c_1}{2d_1\ln2}>\frac{c_2}{d_2\ln2}>\frac{c_2}{2d_2\ln2}$.

Before we start the analysis for the above two cases, the upper
limit and the lower limit of the optimal price $\mu^*$ are given out
in the following two propositions.

\underline{\textbf{Proposition 4.1:}} The upper bound of $\mu^*$ is
$\frac{c_1}{ d_1\ln2}$, i.e., $\mbox{Sup}~(\mu^*)=\frac{c_1}{
d_1\ln2}$.

\begin{proof}
This can be proved by contradiction. Suppose the optimal $\mu^*$ of
P1 satisfies $\mu^*\ge\frac{c_1}{ d_1\ln2}$. Then, it follows that
$\mu^*\ge\frac{c_2}{ d_2\ln2}$ since
$\frac{c_1}{d_1}>\frac{c_2}{d_2}$. According to \eqref{Eq-xi}, we
have $x_1=0$ and $x_2=0$. The resulting revenue for the uploader is
zero. It is easy to see that we can find another pricing strategy
$\mu^\prime$ that satisfies $\mu^\prime<\frac{c_1}{ d_1\ln2}$ and
can generate a revenue larger than zero.  This contradicts with our
presumption. Thus, $\mu^*\ge\frac{c_1}{ d_1\ln2}$ does not hold.
Therefore, $\mu^*$ must be less than $\frac{c_1}{ d_1\ln2}$.
\end{proof}

\underline{\textbf{Proposition 4.2:}} The minimum value for $\mu^*$
is $\frac{c_2}{ 2d_2\ln2}$, i.e., $\mbox{Min}~(\mu^*)=\frac{c_2}{
2d_2\ln2}$.

\begin{proof}
It is observed from \eqref{Eq-xi} that $x_1=d_1$ and $x_2=d_2$ if
$\mu=\frac{c_2}{ 2d_2\ln2}$.  It is clear that $x_1$ and $x_2$ will
not increase if the uploader sets a lower $\mu$, which indicates
that the uploader cannot increase its revenue by setting a price
lower than $\frac{c_2}{ 2d_2\ln2}$. Therefore, the minimum value for
the optimal price $\mu^*$ is $\frac{c_2}{ 2d_2\ln2}$.
\end{proof}

Now, we solve P1 for the two-peer scenario under the above two
cases, respectively. First, we consider Case I: $\frac{c_1}{
d_1\ln2}>\frac{c_2}{d_2\ln2}>\frac{c_1}{2d_1\ln2}>\frac{c_2}{2d_2\ln2}$.
To derive the optimal price $\mu^*$, we consider the following three
possible intervals.

\begin{itemize}
  \item Case I-a: $\mu^*\in\left[\frac{c_2}{d_2\ln2}, \frac{c_1}{
  d_1\ln2}\right)$. In this case, based on  \eqref{Eq-xi}, we have $x_1=\frac{c_1}{\mu\ln2}-d_1$ and $x_2=0$. As a result, P1 is reduced to a convex
  optimization problem, and it can be solved
that $\mu^*=\frac{c_1}{(u_k+d_1)\ln2}$. Using the same approach as
\cite{kangJSAC}, it can be shown that $\mu^*$ is the optimal
solution for P1 if and only if $\frac{c_2}{d_2\ln2}\le
\frac{c_1}{(u_k+d_1)\ln2}< \frac{c_1}{ d_1\ln2}$, i.e.,
$\frac{c_1}{c_2/d_2}-d_1\ge u_k>0$. Thus, it follows that
\begin{align}
\mu^*=\frac{c_1}{(u_k+d_1)\ln2},~\mbox{if}~\frac{c_1}{c_2/d_2}-d_1\ge
u_k>0.
\end{align}
  \item Case I-b:
  $\mu^*\in\left[\frac{c_1}{2d_1\ln2},\frac{c_2}{d_2\ln2}\right)$. In this case, based on  \eqref{Eq-xi}, we have $x_1=\frac{c_1}{\mu\ln2}-d_1$ and
$x_2=\frac{c_2}{\mu\ln2}-d_2$. As a result, P1 becomes a convex
optimization problem, and it can be solved that
$\mu^*=\frac{c_1+c_2}{(u_k+d_1+d_2)\ln2}$. Same as Case I-a, it can
be shown that the obtained $\mu^*$ is the optimal solution for P1 if
and only if $\frac{c_1}{2d_1\ln2}\le
\frac{c_1+c_2}{(u_k+d_1+d_2)\ln2}<\frac{c_2}{d_2\ln2}$, i.e.,
$\frac{c_1+c_2}{c_1/2d_1}-d_1-d_2\ge
u_k>\frac{c_1+c_2}{c_2/d_2}-d_1-d_2$. Thus, it follows that
\begin{align}
\mu^*&=\frac{c_1+c_2}{(u_k+d_1+d_2)\ln2},\nonumber\\&~\mbox{if}~\frac{c_2}{c_1/2d_1}+d_1-d_2\ge
u_k>\frac{c_1}{c_2/d_2}-d_1.
\end{align}
  \item Case I-c: $\mu^*\in\left[\frac{c_2}{2d_2\ln2},\frac{c_1}{2d_1\ln2}\right)$. In this case, based on \eqref{Eq-xi}, we have
$x_1=d_1$ and $x_2=\frac{c_2}{\mu\ln2}-d_2$. Same as the previous
two subcases, P1 is reduced to a convex problem,  and it follows
that $\mu^*=\frac{c_2}{(u_k-d_1+d_2)\ln2}$. It is the optimal
solution for P1 if and only if
$\frac{c_2}{2d_2\ln2}\le\frac{c_2}{(u_k-d_1+d_2)\ln2}<\frac{c_1}{2d_1\ln2}$,
i.e., $d_1+d_2\ge u_k>\frac{c_2}{c_1/2d_1}+d_1-d_2$. Thus, it
follows that
\begin{align}
\mu^*&=\frac{c_2}{(u_k-d_1+d_2)\ln2},\nonumber\\&~\mbox{if}~d_1+d_2\ge
u_k>\frac{c_2}{c_1/2d_1}+d_1-d_2.
\end{align}
\end{itemize}

Based on the above results, the optimal pricing strategy for the
uploader under Case I can be summarized as
\begin{align}\label{Eq-case1mu}
\mu^*\kern-0.5mm=\kern-0.5mm\left\{\kern-1.5mm\begin{array}{ll}
  \frac{c_1}{(u_k+d_1)\ln2}, \kern-1mm& \mbox{if}~\frac{c_1}{c_2/d_2}-d_1\ge
u_k>0, \\
  \frac{c_1+c_2}{(u_k+d_1+d_2)\ln2}, \kern-1mm& \mbox{if}~\frac{c_2}{c_1/2d_1}\kern-0.5mm+\kern-0.5mmd_1\kern-0.5mm-\kern-0.5mmd_2\kern-0.5mm\ge
u_k\kern-0.5mm>\kern-0.5mm\frac{c_1}{c_2/d_2}\kern-0.5mm-\kern-0.5mmd_1, \\
  \frac{c_2}{(u_k-d_1+d_2)\ln2}, \kern-1mm& \mbox{if}~d_1\kern-0.5mm+\kern-0.5mmd_2\ge
u_k>\frac{c_2}{c_1/2d_1}\kern-0.5mm+\kern-0.5mmd_1\kern-0.5mm-\kern-0.5mmd_2.
\end{array}
\right.
\end{align}

\underline{\textbf{Remark:}} It is observed from \eqref{Eq-case1mu}
that the optimal price can be divided into three regions based on
the uploader's available bandwidth. 
Based on the demand of the peers and the supply of the uploader, the
three regions are named as \emph{insufficient region}, \emph{balance
region}, and \emph{sufficient region}. In the insufficient region,
the uploader's bandwidth is not enough to support all the peers. In
this region, at least one peer will be not assigned any bandwidth.
Peers are excluded from the game based on their $\frac{c_i}{d_i}$
values. Peers with low values are rejected first. For instance, in
Case I-a, peer $2$ is rejected from the game and only peer $1$
remains in the game. In the balance region, the uploader will
allocate bandwidth to each peer. However, none of the peers can
download at its maximum bandwidth $d_i$. In this region, the
uploader allocates its limited bandwidth to the peers proportional
to their $\frac{c_i}{d_i}$ values. 
In the sufficient region, the uploader's bandwidth is able to
support both the peers, and at least one of them can download at its
maximum download bandwidth. The peer with the largest
$\frac{c_i}{d_i}$ will be the first peer that can download at its
maximum download bandwidth. When the bandwidth is sufficiently
large, both peers can download at their maximum download bandwidth.
~~~~~~~~~~~~~~~~~~~~~~~~~~~~~~~~~~~~~~~~~~~~~~~~~~~~~~~$\blacksquare$

Now, we consider Case II:
$\frac{c_1}{d_1\ln2}>\frac{c_1}{2d_1\ln2}>\frac{c_2}{d_2\ln2}>\frac{c_2}{2d_2\ln2}$.
Similar as Case I, we consider different intervals to find the
optimal price $\mu^*$ in each interval.
\begin{itemize}
  \item Case II-a: $\mu^*\in\left[\frac{c_1}{2d_1\ln2},
  \frac{c_1}{d_1\ln2}\right)$. In this case, based on  \eqref{Eq-xi}, we have $x_1=\frac{c_1}{\mu\ln2}-d_1$ and $x_2=0$. Same as Case I-a, P1 becomes a convex
  optimization problem, and it can be solved
that $\mu^*=\frac{c_1}{(u_k+d_1)\ln2}$. Using the same approach as
\cite{kangJSAC}, it can be shown that $\mu^*$ is the optimal
solution for P1 if and only if $\frac{c_1}{2d_1\ln2}\le
\frac{c_1}{(u_k+d_1)\ln2}<\frac{c_1}{ d_1\ln2}$, i.e., $d_1\ge
u_k>0$. Thus, it follows that
\begin{align}\label{Eq-case2a}
\mu^*=\frac{c_1}{(u_k+d_1)\ln2},~\mbox{if}~d_1\ge u_k>0.
\end{align}
  \item Case II-b: $\mu^*\in\left[\frac{c_2}{d_2\ln2},
  \frac{c_1}{2d_1\ln2}\right)$. In this case, we show that the maximum possible utility for the uploader is
  lower than that obtained in Case II-a, and hence the optimal price will never lie in this range. Based on  \eqref{Eq-xi}, we
  have $x_1=d_1$ and $x_2=0$. P1 is thus reduced to finding the maximum value of $\mu d_1$, and is valid only when $d_1\le
  u_k$. Since $\mu \in\left[\frac{c_2}{d_2\ln2},
  \frac{c_1}{2d_1\ln2}\right)$, the upper bound of $\mu d_1$ is
  $\frac{c_1}{2\ln2}$.
  However, when $d_1 \le u_k$, it is observed from Case II-a that the maximum revenue
  is $\mu^* \left(\frac{c_1}{\mu^*\ln2}-d_1\right)$, where $\mu^*$
  is given by
  \eqref{Eq-case2a}. Thus,  $\mu^* \left(\frac{c_1}{\mu^*\ln2}-d_1\right)$ can be computed as $\frac{c_1u_k}{(u_k+d_1)\ln2}$,
  which is larger than $\frac{c_1}{2\ln2}$ when $d_1 \le u_k$. Thus, $\mu^*$ should not
  lie in this range.
  \item Case II-c: $\mu^*\in\left[\frac{c_2}{2d_2\ln2},
  \frac{c_2}{d_2\ln2}\right)$. In this case, based on  \eqref{Eq-xi}, we have $x_1=d_1$ and
  $x_2=\frac{c_2}{\mu\ln2}-d_2$. It follows
that $\mu^*=\frac{c_2}{(u_k-d_1+d_2)\ln2}$. It is the optimal
solution for P1 if and only if $\frac{c_2}{2d_2\ln2}\le
\frac{c_2}{(u_k-d_1+d_2)\ln2}<\frac{c_2}{d_2\ln2}$, i.e.,
$d_1+d_2\ge u_k>d_1$.
\begin{align}
\mu^*=\frac{c_2}{(u_k-d_1+d_2)\ln2},~\mbox{if}~d_1+d_2\ge u_k>d_1.
\end{align}
\end{itemize}

Based on the above results, the optimal pricing strategy for the
uploader under Case II can be summarized as
\begin{align}\label{Eq-case2mu}
\mu^*\kern-0.5mm=\kern-0.5mm\left\{\kern-1.5mm\begin{array}{ll}
  \frac{c_1}{(u_k+d_1)\ln2}, & \mbox{if}~d_1\ge u_k>0, \\
  \frac{c_2}{(u_k-d_1+d_2)\ln2}, & \mbox{if}~d_1+d_2\ge u_k>d_1.
\end{array}
\right.
\end{align}

\underline{\textbf{Remark:}} It is observed from \eqref{Eq-case2mu}
that the optimal price obtained under Case II can be divided into
two regions based on the uploader's available bandwidth. We refer to
these two regions as \emph{insufficient region} and \emph{sufficient
region}. In the insufficient region, the uploader will only accept
the request from peer $1$, which is the peer with high
$\frac{c_i}{d_i}$ value. In the sufficient region, the uploader will
allocate peer $1$ its maximum download bandwidth. It is observed
that the price strategy will allocate bandwidth to peer $2$ only
when peer $1$ is allocated its full download bandwidth $d_1$. This
is quite different from the scenario in Case I. This phenomenon
happens due to the fact that the peer $1$'s $\frac{c_i}{d_i}$ value
is much larger than that of peer $2$.
~~~~~~~~~~~~~~~~~~~~~~~~~~~~~~~~~~~ $\blacksquare$

\begin{figure}[t]
        \centering
        \includegraphics*[width=8.5cm]{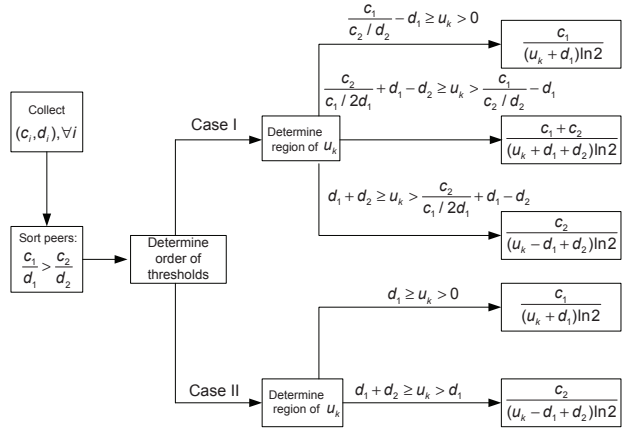}
        \caption{Procedure of finding the optimal price for the two-peer scenario}
        \label{Fig-IlluOPPriceAllo}
\end{figure}
In summary, the procedure to find the optimal price for the two-peer
scenario is given in Fig. \ref{Fig-IlluOPPriceAllo}. It is not
difficult to observe that the optimal price is determined by the
following two factors:
\begin{itemize}
  \item The order of the downloaders' thresholds.
  \item The uploader's available upload bandwidth.
\end{itemize}

Once these two factors are determined, the optimal price can be
easily obtained. From an economic perspective, the order of the
downloaders' thresholds actually reflect the demand and the
purchasing power (i.e., the available accumulated credits) of the
downloaders. The uploader's upload bandwidth reflects the market
supply. The price of the goods is determined by the relationship
between the supply and demand.

Another key result observed from the above solutions is that the
optimal price $\mu^*$ is always obtained when \eqref{P1-con} holds
with equality, i.e., $\sum_{i\in S_k} f_i(\mu^*) =u_k$. This
observation is very important, and plays a significant role in
determining the Stackelberg equilibrium of the proposed game, which
will be discussed later in Subsection 5.3.

\subsubsection{Multi-peer Scenario}
For the multi-peer scenario, there are more cases, and the number of
cases increases with the increase of the number of peers. Thus, in
general, we are not able to obtain a closed-form solution for the
multi-peer scenario. However, once the order of the peers'
thresholds is determined, a closed-form solution can be obtained.
For the purpose of illustration, we derive the closed-form solution
for P1 when the thresholds of the peers satisfy the following order
$\frac{c_1}{d_1}>\cdots>\frac{c_{|S_k|}}{d_{|S_k|}}>\frac{c_1}{2d_1}>\cdots>\frac{c_{|S_k|}}{2d_{|S_k|}}$,
where $|\cdot|$ denotes the cardinality of a set.

To avoid trivial solutions, we assume that $\sum_{i \in\mathcal
{S}_k}d_i>u_k$ in the following analysis. Due to the complexity of
$f_i(\mu)$, P1 is difficult to solve directly. Therefore, to solve
P1, we first consider the following problem
\begin{align}
\mbox{P2:}~~\max_{\mu> 0}~&\sum_{i\in S_k} \mu \left(\frac{c_i}{\mu\ln2}-d_i\right),\\
\mbox{s.t.}~~& \sum_{i\in S_k} \left(\frac{c_i}{\mu\ln2}-d_i\right)
\le u_k.
\end{align}

This problem is a convex optimization problem. Therefore, this
problem can be solved by standard convex optimization techniques.
Details are omitted here for brevity. The optimal price $\mu$ for P2
can be obtained as follows,
\begin{align}\label{Eq-mui-op}
\mu=\frac{\sum_{i\in S_k}c_i}{\left(u_k+\sum_{i\in S_k}
d_i\right)\ln2}.
\end{align}

Now, we relate the optimal solution of P2 to that of P1 in the
following proposition.

\underline{\textbf{Proposition 4.3:}} The price $\mu$ given in
\eqref{Eq-mui-op} is the optimal solution of P1 if and only if
\begin{align}\label{Eq-26} \frac{\sum_{i\in S_k}c_i}{\min_i \kern-0.5mm
\frac{c_i}{d_i}}-\kern-1mm\sum_{i\in S_k}\kern-0.5mm
d_i<u_k<\frac{2\sum_{i\in S_k}c_i}{\max_i \kern-0.5mm
\frac{c_i}{d_i}}-\kern-1mm\sum_{i\in S_k}\kern-0.5mm d_i,
\end{align} when $\min_i \frac{c_i}{d_i}\ge \max_i
\frac{c_i}{2d_i}$.

\begin{proof}
This proof consists of the following two parts.

Part 1: Sufficiency. The optimal price $\mu^*$ given by
\eqref{Eq-mui-op} is the optimal solution of P1 if
$\frac{c_i}{2d_i\ln2}\le\mu<\frac{c_i}{d_i\ln2}, \forall~i$.
Substituting \eqref{Eq-mui-op} into these inequalities yields $
\frac{\sum_{i\in
S_k}\kern-0.5mmc_i}{\frac{c_i}{d_i}}-\kern-0.5mm\sum_{i\in S_k}
d_i<u_k\le\frac{2\sum_{i\in
S_k}\kern-0.5mmc_i}{\frac{c_i}{d_i}}-\kern-0.5mm\sum_{i\in S_k}d_i,
\forall~i. $ Thus, when $\min_i \frac{c_i}{d_i}\ge \max_i
\frac{c_i}{2d_i}$, these inequalities can be compactly written as
\begin{align}\label{Eq-003}
\frac{\sum_{i\in S_k}c_i}{\min_i \kern-0.5mm
\frac{c_i}{d_i}}-\kern-1mm\sum_{i\in S_k}\kern-0.5mm
d_i<u_k\le\frac{2\sum_{i\in S_k}c_i}{\max_i \kern-0.5mm
\frac{c_i}{d_i}}-\kern-1mm\sum_{i\in S_k}\kern-0.5mm d_i.
\end{align}
The ``if'' part is thus proved. Next, we consider the ``only if''
part, which is proved by contradiction as follows.

Part 2: Necessity. For the ease of exposition, we assume that the
peers are sorted by the following order:
$\frac{c_1}{d_1}>\cdots>\frac{c_{|S_k|}}{d_{|S_k|}}>\frac{c_1}{2d_1}>\cdots>\frac{c_{|S_k|}}{2d_{|S_k|}}.$ 
In order to prove the necessity, we suppose that the price $\mu^*$
given in \eqref{Eq-mui-op} is optimal even if the inequality given
in \eqref{Eq-26} does not hold. We consider on possible region for
$\mu^*$ below, and a similar proof applies to the other regions.
Suppose $u_k$ satisfies the following inequality
\begin{align}\label{Eq-001}
\frac{\sum_{i=1}^{|S_k|-1}c_i}{\frac{c_{|S_k|-1}}{d_{|S_k|-1}}}-\kern-1mm\sum_{i=1}^{|S_k|-1}\kern-0.5mm
d_i<u_k\le\frac{\sum_{i=1}^{|S_k|}c_i}{\frac{c_{|S_k|}}{d_{|S_k|}}}-\kern-1mm\sum_{i=1}^{|S_k|}\kern-0.5mm
d_i,
\end{align}
and $\mu^*$ given by \eqref{Eq-mui-op} is still optimal when
\eqref{Eq-001} holds. Since
$u_k\le{\sum_{i=1}^{|S_k|}c_i}/{\frac{c_{|S_k|}}{d_{|S_k|}}}-\kern-1mm\sum_{i=1}^{|S_k|}\kern-0.5mm
d_i$, it follows that
$\frac{c_{|S_k|}}{d_{|S_k|}}\le\frac{\sum_{i=1}^{|S_k|}c_i}{u_k+\sum_{i=1}^{|S_k|}\kern-0.5mm
d_i}$. Then, according to \eqref{Eq-mui-op}, we have that
$\mu^*\ge\frac{c_{|S_k|}}{d_{|S_k|}}$. Then, it follows from
\eqref{Eq-xi} that $x_{|S_k|}=0$. This indicates that the peer with
the smallest  $\frac{c_{i}}{d_{i}}$ will be excluded from the game
under the above condition.

Then, it follows that $\mu^*$ must be the optimal solution of P1
with $|S_k|-1$ peers, which is given as follows
\begin{align}
\max_{\mu> 0}~\sum_{i=1}^{|S_k|-1} \mu f_i(\mu), ~~\mbox{s.t.}~
\sum_{i=1}^{|S_k|-1} f_i(\mu) \le u_k.
\end{align}

Thus, under the condition given by \eqref{Eq-001}, using the same
way as the proof of the previous ``if'' part, it can be shown that
the optimal solution for this problem is given by
\begin{align}\label{Eq-002}
\tilde{\mu}^*=\frac{\sum_{i=1}^{|S_k|-1}c_i}{\left(u_k+\sum_{i=1}^{|S_k|-1}
d_i\right)\ln2}.
\end{align}

It is easy to observe that the optimal price $\tilde{\mu}^*$ given
in \eqref{Eq-002} for the above problem is different from $\mu^*$
given by \eqref{Eq-mui-op}. Thus, this contradicts with our
presumption that $\mu^*$ is optimal for P1 with $u_k$ satisfying
\eqref{Eq-001}. Using the same method, we can prove that $\mu^*$ is
not the optimal for P1 for other regions. Therefore, the
interference vector $\mu^*$ given by \eqref{Eq-mui-op} is the
optimal solution of P1 only if $u_k$ satisfying \eqref{Eq-003}. The
``only if'' part thus follows.

By combining the proofs of both the ``\emph{sufficiency}'' and
``\emph{necessity}'' parts, Proposition 4.3 is thus proved.
\end{proof}

With the results obtained above, we can solve a series of similar
sub-problems of P1. Then, combing these obtained results by the same
approach as the two-peer scenario, we can obtain the following
theorem.

\underline{\textbf{Theorem 4.2:}} When the thresholds of the peers
satisfy $h_1>\cdots>h_{|S_k|}>h_1/2>\cdots>h_{|S_k|}$/2, where
$h_i\triangleq c_i/d_i$, the optimal price $\mu^*$ for P1 is then
given by
\begin{align}\label{Eq-mui-opstar}
\mu^*=\left\{\begin{array}{ll}
p_{|S_k|}, &\mbox{if}~R_{|S_k|}\ge u_k>R_{|S_k|-1}  \\
                            \vdots&\vdots\\
                            p_{2}, &\mbox{if}~R_2\ge u_k>R_1 \\
                            q_{|S_k|}, &\mbox{if}~R_1\ge u_k>T_{|S_k|}  \\
                            q_{|S_k|-1}, &\mbox{if}~T_{|S_k|}\ge u_k>T_{|S_k|-1}  \\
\vdots&\vdots
                            \\q_{1},
                            &\mbox{if}~T_{2}\ge u_k>T_{1}
                          \end{array}
\right.,
\end{align}
where $q_K=\frac{\sum_{i=1}^K c_i}{\left(u_k+\sum_{i=1}^K
d_i\right)\ln2}$,
$T_K=\frac{\sum_{i=1}^Kc_i}{h_K}-\kern-1mm\sum_{i=1}^K\kern-0.5mm
d_i$, $p_K=\frac{\sum_{i=K}^{|S_k|}c_i}{ \left(u_k-\sum_{i=1}^{K-1}
d_i+\sum_{i=K}^{|S_k|} d_i\right)\ln2}$ and
$R_K=\frac{2\sum_{i=K}^{|S_k|}c_i}{h_K}+\sum_{i=1}^{K-1}
d_i-\kern-1mm\sum_{i=K}^{|S_k|}\kern-0.5mm d_i$.


For other cases of the multi-peer scenario, closed-form solutions
can also be obtained in the same way. In general, the optimal
pricing strategy for the multi-peer scenario can be obtained by the
same procedure illustrated in Fig. \ref{Fig-IlluOPPriceAllo}. For
the same type of peer (i.e., the same $d$), the optimal pricing
scheme tends to allocate more bandwidth to peers with higher
contribution (i.e., more credits $c$). This indicates that the
obtained pricing strategy for the multi-peer scenario can provide a
strong incentive for peers to cooperate with each other. It is also
observed that the optimal price $\mu^*$ is always obtained when
\eqref{P1-con} holds with equality, i.e., $\sum_{i\in S_k}
f_i(\mu^*) =u_k$.

\subsection{Stackelberg Equilibrium of the Proposed Game}
In this subsection, we investigate the SE for the proposed
Stackelberg game, and show the SE is unique and Pareto-optimal when
$u_k$ is given.

With the optimal solution obtained in Theorem 4.1 and 4.2, the SE
for the proposed Stackelberg game is given as follows.

\underline{\textbf{Theorem 4.3:}} The SE for the Stackelberg game
formulated by the uploading game and the downloading game is
$(\boldsymbol{x}^*, \mu^*)$, where $\boldsymbol{x}^*$ is given by
\eqref{Eq-xi}, and $\mu^*$ is the optimal solution of P1.

\begin{proof}
Since $\mu^*$ is the optimal solution of P1, we have
$U^{up}\left(\mu^*,\boldsymbol{x}\right)\ge
U^{up}\left(\mu,\boldsymbol{x}\right)$ for any given vector
$\boldsymbol{x}$. Thus, it follows that
$U^{up}\left(\mu^*,\boldsymbol{x}^*\right)\ge
U^{up}\left(\mu,\boldsymbol{x}^*\right)$, where $\boldsymbol{x}^*$
is given by \eqref{Eq-xi}. Similarly, since  $\boldsymbol{x}^*$ is
the optimal solution for the downloading game, we have
$U^{down}_i\left(x_i^*,\mu\right)\ge U^{down}_i\left(x_i,\mu\right),
\forall i,$ for any given price $\mu$. Thus, it follows that
$U^{down}_i\left(x_i^*,\mu^*\right)\ge
U^{down}_i\left(x_i,\mu^*\right), \forall i$. Then, combining the
above two facts, according to the definition of SE given in
Definition 3.1, $(\boldsymbol{x}^*, \mu^*)$ is the SE for the
proposed Stackelberg game.
\end{proof}

Now, we show that the SE is unique and Pareto-optimal when $u_k$ is
given. 
%

\underline{\textbf{Theorem 4.4:}} The SE for the proposed
Stackelberg game is unique and Pareto-optimal for a given $u_k$.

\begin{proof}
First, we show that the SE for the proposed Stackelberg game is
unique for a given $u_k$. As pointed out in the previous subsection,
the optimal pricing strategy is unique when the order of peers'
thresholds and $u_k$ are given. The order of peers' thresholds is
determined by the values of $c_i$ and $d_i$, $\forall i$, which are
fixed during each implementation of the Stackelberg game. Thus, it
is clear that the optimal price $\mu^*$ is unique for a given $u_k$.
On the other hand, it is observed from \eqref{Eq-xi} that the
download bandwidth for each peer is unique under a given $\mu$.
Thus, it is obvious that the SE for the proposed game is unique
under a given $u_k$.

Now, we show that the SE is Pareto-optimal for a given $u_k$. Given
an initial resource allocation scheme among a group of peers, a
change to a different allocation scheme that makes at least one peer
better off without making any other peers worse off is called a
Pareto improvement. An allocation scheme is defined as
"Pareto-optimal" when no further Pareto improvements can be made. In
other words, in a Pareto-optimal equilibrium, no one can be made
better off without making at least one individual worse off. It is
observed that the optimal $\mu^*$
always satisfies $\sum_{i\in S_k} f_i(\mu^*)=u_k$. 
Thus, increasing one peer's (e.g., peer 1) bandwidth allocation will
inevitably decrease another peer's (e.g. peer 2) bandwidth
allocation. This makes peer 2's bandwidth allocation deviates from
its optimal bandwidth allocation, and consequentially decreasing its
utility. Thus, no peer can be made better off without making some
other peer worse off, and the SE is Pareto-optimal.
\end{proof}

\section{Implementation of the Stackelberg Game in P2P Streaming
Networks} \label{Sec-implementation} In previous section, we have
solved the proposed Stackelberg game and obtained its SE. In this
section, we investigate how to implement the proposed game in P2P
networks in detail. Two implementation methods referred to as
\emph{direct implementation} and \emph{bargaining implementation}
are proposed and investigated as below.

\subsection{Direct Implementation}
Direct implementation is strictly based on the obtained results
given in Section \ref{Sec-Op solution}. It is a one-round
implementation with four stages, which are described as follows.

\begin{itemize}
\item \textbf{Stage 1:} All the peers requesting data from the
uploader send their contribution values $c_i$ and download bandwidth
limits $d_i$ to the uploader.

\item \textbf{Stage 2:} Having received the data requests together
with the information of $c_i$ and $d_i$ from all the peers, the
uploader first sorts all the peer in the order
$\frac{c_1}{d_1}>\cdots>\frac{c_|\mathcal {S}_k|}{d_|\mathcal
{S}_k|}$, and determines the order of peers' thresholds. Then, the
uploader computes the optimal price $\mu^*$ using the same approach
as illustrated in Fig. \ref{Fig-IlluOPPriceAllo}, and broadcasts the
optimal price $\mu^*$ to all the peers.

\item \textbf{Stage 3:} Based on the received price, each peer
computes its optimal download bandwidth $x^*_i$ based on
\eqref{Eq-xi}, and sends the calculated results to the uploader.

\item \textbf{Stage 4:}  The uploader allocates the bandwidth
based on $x^*_i, \forall i \in \mathcal {S}_k$ and starts streaming.
\end{itemize}

\subsection{Bargaining Implementation}
\begin{figure}[t]
        \centering
        \includegraphics*[width=8cm]{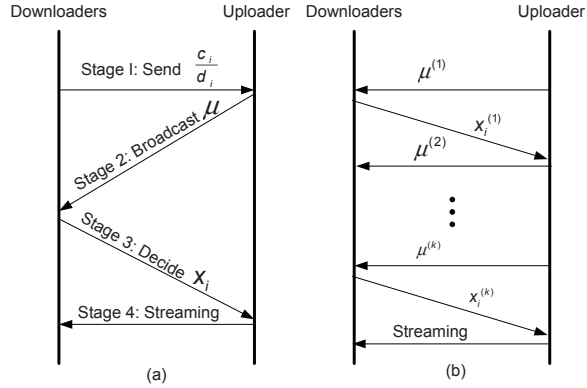}
        \caption{Implementation in P2P Streaming Networks: (a) Direct implementation (b) Bargaining implementation}
        \label{Fig-BargainProcess}
\end{figure}
In this subsection, we propose the bargaining implementation for the
proposed Stackelberg game based on the characteristics of P1. It can
be shown that P1 is equivalent to the following problem
\begin{align}
\mbox{P3:}~~\max_{\mu> 0}~&~~\mu,\\
\mbox{s.t.}~~& \sum_{i\in S_k} f_i(\mu) = u_k.
\end{align}
Then, we propose the bargaining implementation based on the
following two facts: (i). $f_i(\mu)$ is a decreasing function of
$\mu$, which can be observed from \eqref{Eq-xi}. (ii). The upper
limit of $\mu^*$ is $\max_i \frac{c_i}{d_i\ln2}$. This can be proved
using the same approach as Proposition 4.1.

\begin{itemize}
\item \textbf{Stage 1:} The uploader sets an initial price $\mu$ (where $\mu \ge \max_i \frac{c_i}{d_i\ln2}$), and
broadcasts it to all the downloaders.

\item \textbf{Stage 2:} Each downloader computes its optimal download bandwidth $x_i$ based on
\eqref{Eq-xi} for the given $\mu$, and send back $x_i$ to the
uploader.

\item \textbf{Stage 3:} Having received $x_i$ from all the peers, the
uploader computes the total demand $\sum_{i\in \mathcal {S}_k} x_i$,
and compares the total demand with its upload bandwidth $u_k$.
Assume that $\epsilon$ is a small positive constant that controls
the algorithm accuracy. If $\sum_{i\in \mathcal {S}_k}
x_i<u_k-\epsilon$, the uploader decreases the bandwidth price by
$\Delta \mu$, where $\Delta \mu$ is a small step size. After that,
the uploader broadcasts the new price to all the downloaders.

\item \textbf{Stage 4:} Stage 2 and Stage 3 are
repeated until $|\sum_{i\in \mathcal {S}_k} x_i-u_k|<\epsilon$.
Then, the uploader starts streaming.
\end{itemize}

The convergence of the bargaining algorithm is guaranteed by the
following facts: (i). The optimal price $\mu^*$ is always obtained
when the upload bandwidth of the uploader is fully allocated, i.e.,
$\sum_{i\in \mathcal {S}_k} x_i=u_k$. (ii). $f_i(\mu)$ is a
decreasing function of $\mu$.  (iii). The SE for the proposed
Stackelberg game is unique and Pareto-optimal for a given $u_k$.


\subsection{Direct Implementation Vs. Bargaining Implementation}
In this subsection, we analyze and compare the difference between
these two kinds of implementation schemes.

It is not difficult to observe that the direct implementation is
time-saving, since it only needs one-round to determine the optimal
bandwidth price and the optimal download bandwidth of the peers. In
contrast, the bargaining implementation requires much more time.
This is due to the fact that the uploader and the downloaders have
to go through a multi-round bargaining process to finally reach the
equilibrium. Thus, for delay-sensitive service, such as P2P
multimedia streaming, direct implementation is preferred.

Another difference between these two implementation schemes is the
requirement on the computing power of the uploader. It is observed
that direct implementation requires the uploader to directly compute
the optimal price based on the procedure given in Section
\ref{Sec-OPS}, which is a complex procedure involving a lot of
cases. Thus, it has a high requirement on the computing power of the
uploader. In contrast, the bargaining implementation greatly
relieves the computation burden on the uploader. The uploader only
needs to compare the total demand with its upload bandwidth, which
requires much less computing power. Thus, the bargaining
implementation should be preferred by the handheld mobile devices
with less computing power.

It is worthy pointing out that no matter which implementation scheme
is employed,  the same Stackelberg equilibrium  results for the same
set data. This is due to the fact that the SE is unique which is
proved in Theorem 4.4.

\section{Dealing with Dynamics of P2P Streaming
Networks}\label{Sec-DynamicsAnalysis} P2P networks are dynamic in
nature. Peers may leave or join the network at any time. How the
equilibrium changes when peers leave or join the network is of great
importance to the study of a dynamic network. Thus, in this
subsection, we investigate whether the equilibrium will change and
how it will change under these situations.

When a peer joins the network, it is given a certain number of
credits. The initial credits for each peer can be the same (e.g.,
$100$ credits for each peer) or different (e.g., $c_i$ for peer
$i$). The credits of a peer is updated after each transaction. One
transaction means that a downloading peer finished its downloading
from a uploader. After one transaction of downloader $i$, its
credits $c_i$ is updated by $c_i=c_i-\mu*x_i$, and the credits of
the uploader $j$ is updated by $c_j=c_j+\mu*x_i$. If multiple
downloader finish their downloading at the same time, the uploader
updates its credits by collecting credits from them together.

To facilitate the analysis, we assume that there are $N$ downloading
peers and $1$ uploading peer at the original SE. The original SE is
denoted by $(\mu^*, \boldsymbol{x}^*)$, where $\boldsymbol{x}^*$ is
the optimal bandwidth allocation vector for the downloading peers at
the SE. The new SE after peers leaving or joining the network is
denoted by $(\tilde{\mu}^*, \tilde{\boldsymbol{x}}^*)$. Besides, we
assume that the information that a downloader leaves or joins the
network is only available to the uploader itself, and it will not
share the information with other downloaders.

\subsection{Peers Leaving the P2P Streaming
Network}\label{Subsec-peersleave}
When a peer $j$ leaves a P2P
streaming network, the SE changes only when
\begin{align}\label{eq-34}U^{up}(\tilde{\mu}^*, \tilde{\boldsymbol{x}}^*)\ge
U^{up}\left(\mu^*,\boldsymbol{x}^*\right)-U^{up}\left(\mu^*,x_j\right),\end{align}
where $U^{up}(\tilde{\mu}^*, \tilde{\boldsymbol{x}}^*)$ denotes the
utility of the uploader at the new SE,
$U^{up}\left(\mu^*,\boldsymbol{x}^*\right)$ denotes the utility of
the uploader at the original SE, and $U^{up}\left(\mu^*,x_j\right)$
denotes peer $j$'s contribution to the uploader's utility at the
original SE.


For the problem considered in this paper, the inequality
\eqref{eq-34} always holds.  Thus, when a downloading peer leaves
the network, the best strategy is to re-implement the Stackelberg
game with the remaining $N-1$ peers.

\subsection{Peers Joining the P2P Streaming
Network}\label{Subsec-peersjoin} When a peer joins the network, the
SE changes only when
\begin{align}\label{eq-35}U^{up}(\tilde{\mu}^*,
\tilde{\boldsymbol{x}}^*)>
U^{up}\left(\mu^*,\boldsymbol{x}^*\right).\end{align} When a peer
joins the network, the number of competing peers increases. As the
competition between downloading peers becomes fiercer, the uploading
peer has the incentive to increase the price of the resource to
increase its revenue. It is worth pointing out that \eqref{eq-35}
does not always hold. For example, if the $c_i/d_i$ value of the
joining peer is very small, this peer will be rejected, and the SE
will be sustained.

A simple way to re-attain the equilibrium is to completely
re-implement the Stackelberg game again with the $N+1$ peers. This
method is guaranteed to reach a new SE which is unique and
Pareto-optimal. From the uploader's perspective, this method is
beneficial since the new SE will never decrease its utility. The
uploader
 need only recompute the price $\mu^*$ of the resource, taking the
new peer's $c_i$ and $d_i$ into consideration, and broadcasts the
calculated $\mu_{new}^*$ to all the $N+1$ peers. Though this method
is good for the uploader, it sacrifices the existing downloaders'
interests. This is due to the fact that $\sum_{i\in \mathcal
{S}_k}x_i=u_k$ always holds at the equilibrium. If a new peer joins
and is allocated a certain amount of download bandwidth, some of the
existing peers' download bandwidth must decrease. For some peers,
even though their download bandwidth may not decrease, their utility
decreases due to the increase of the resource price.
\section{Performance Evaluation} \label{NumericalResults}
In this section, several numerical examples are provided to evaluate
the performance of the proposed incentive resource allocation
scheme. It is shown that the proposed resource allocation scheme can
provide strong incentives for peers to contribute to the P2P
network.

\subsection{Example 1: Peers with the same connection type but different contribution values}
\begin{figure}[t]
        \centering
        \includegraphics*[width=8cm]{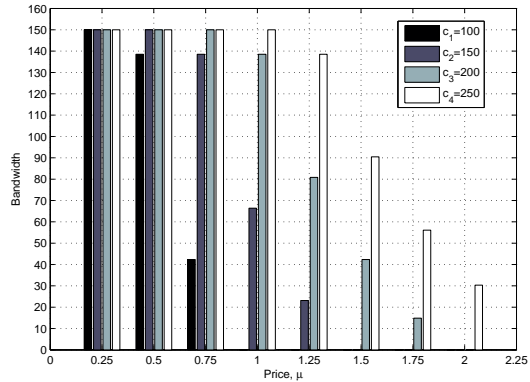}
        \caption{Bandwidth allocation vs. Price under same connection types}
        \label{Fig-P2PIncenFig1}
\end{figure}
In this example, we assume that there are four peers requesting data
chunks from the uploader $k$. The connection types of all the
requesting peers are assumed to be the same, and their maximum
download bandwidth are $d_1=d_2=d_3=d_4=150$. The contribution
values of these requesting peers at the current time are
$[c_1,c_2,c_3,c_4]=[100,150,200,250]$. The bandwidth assignments for
the four downloading peers under different prices are shown in Fig.
\ref{Fig-P2PIncenFig1}.

It is observed from Fig. \ref{Fig-P2PIncenFig1} that the bandwidth
assigned to each peer decreases with the increase of the price. For
the same price, although all the peers have the same connection
type, peers with larger contribution values are assigned higher
bandwidth under our resource allocation strategy. This illustrates
that the proposed resource allocation strategy can provide
differentiated service to peers with different contribution, and
thus encourage peers to contribute to the network.

\subsection{Example 2: Peers with the same contribution value but different connection types}
\begin{figure}[t]
        \centering
        \includegraphics*[width=8cm]{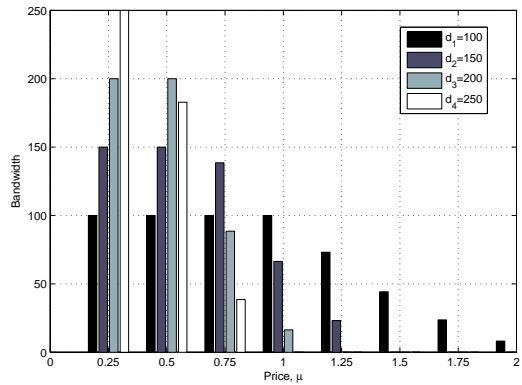}
        \caption{Bandwidth allocation vs. Price under same contribution values}
        \label{Fig-P2PIncenFig4}
\end{figure}
In this example, we assume that there are four peers requesting data
chunks from the uploader $k$. We assume the contribution values of
the requesting peers are the same, and are given by
$[c_1,c_2,c_3,c_4]=[150,150,150,150]$. The connection types of the
requesting peers are assumed to be different, and are given by
$[d_1,d_2,d_3,d_4]=[100,150,200,250]$. The bandwidth assignments for
this setup are shown in Fig. \ref{Fig-P2PIncenFig4}.

It is observed from Fig. \ref{Fig-P2PIncenFig4} that when the price
is low, every peer can download at its maximum download bandwidth.
The bandwidth assigned to each peer decreases with the increase of
the price. It is also observed that our resource allocation scheme
biases toward peers with smaller download capacities. This is as
expected. Intuitively, given the same unit of bandwidth resource, a
peer with a smaller download capacity achieves a higher performance
satisfaction factor than a peer with a larger download capacity.

\subsection{Example 3: Relationship between bandwidth allocation and the available upload bandwidth}
\begin{figure}[t]
        \centering
        \includegraphics*[width=8cm]{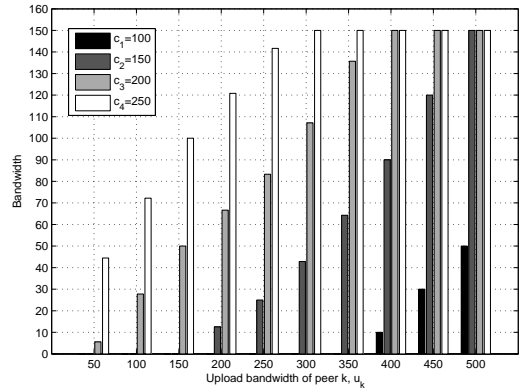}
        \caption{Bandwidth allocation vs. $u_k$}
        \label{Fig-P2PIncenFig2}
\end{figure}

In this example, we assume that there are four peers requesting data
chunks from the uploader $k$. The connection types of all the
requesting peers are assumed to be the same, and their maximum
download bandwidths are $d_1=d_2=d_3=d_4=150$. The contribution
values of these requesting peers at the current time are
$[c_1,c_2,c_3,c_4]=[100,150,200,250]$. The bandwidth assignments for
the four downloading peers under $u_k$ are shown in Fig.
\ref{Fig-P2PIncenFig2}. It is observed from Fig.
\ref{Fig-P2PIncenFig2} that our resource allocation scheme gives a
higher priority to peers with higher contribution in bandwidth
assignment. When the available upload bandwidth $u_k$ is small, the
uploader will reject the request from the peers with low
contribution, and provide the limited resource to the peers with
high contribution. When the available upload bandwidth $u_k$ is
large, the uploader will try to meet every peer's request. However,
peers with higher contribution values are given a higher priority in
obtaining the bandwidth. It is also observed that with the
increasing of $u_k$, the bandwidth assigned for each peer increases.
This is due to the fact that the uploader's utility is maximized
only when it contributes all its available upload bandwidth.

\subsection{Example 4: Join of Competing Peers}
\begin{figure}[t]
        \centering
        \includegraphics*[width=8cm]{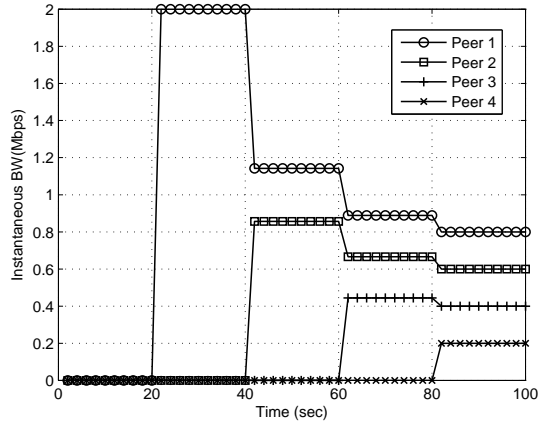}
        \caption{Instantaneous Bandwidth Allocations: Join of Competing Peers}
        \label{Fig-P2PIncenJournalFig1}
\end{figure}

In this example, for the purpose of comparison, we use the same
system setup and simulation parameters as \cite{RMa_200610}. We
assume the uploader's available bandwidth $u_k$ is $2$ Mb/s. There
are four competing peers requesting data chunks from the uploader
$k$. The connection types of the requesting peers are assumed to be
different, and are given by $[d_1,d_2,d_3,d_4]=[2,1.5,1,0.5]$ (in
Mb/s). The arrival times of these peers are $t=20, 40, 60$ and
$80$s, respectively. The contribution values of these requesting
peers are assumed to be $[c_1,c_2,c_3,c_4]=[400,300,200,100]$. The
results are obtained using the bargaining implementation. The
initial value for $\mu$ is $\mu^{(0)}=\max_i
\frac{c_i}{d_i\ln2}=288.539$. The step size for $\mu$ is chosen as
$0.01$, and $\epsilon$ is chosen as $0.001$.

It is observed from Fig. \ref{Fig-P2PIncenJournalFig1} that the
equilibrium of the game changes whenever a new competing peer joins
the game. The bandwidth allocation for the existing peers decrease
due to the newcomer. This is in accordance with our analysis given
in Section \ref{Subsec-peersjoin}. It is also observed that the
uploader assigns all its bandwidth without reservation  at each new
equilibrium. Besides, at each equilibrium, the bandwidth allocation
is proportional to the contribution value of each peer. This
indicates that the proposed incentive mechanism is adaptive to the
dynamics of the P2P network, and can always provide differentiated
service to peers with different contribution values. It is also
observed that the proposed scheme can achieve the same performance
as that of \cite{RMa_200610}.

\subsection{Example 5: Leave of Competing Peers}
\begin{figure}[t]
        \centering
        \includegraphics*[width=8cm]{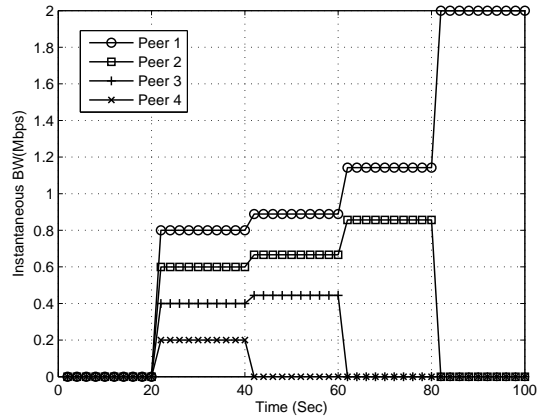}
        \caption{Instantaneous Bandwidth Allocations: Leave of Competing Peers}
        \label{Fig-P2PIncenJournalFig2}
\end{figure}

In this example, we consider an opposite scenario of Example 4. We
consider the scenario that peers leave the system one by one. For
the convenience of analysis, we use the same system setup and
simulation parameters as example 4. We assume that the four peers
join the network at $t=20$s, and they leave the network one by one.
The leave times of peer $4$, $3$, $2$ are $t=40,60,80$s,
respectively. 

%

From Fig. \ref{Fig-P2PIncenJournalFig2}, it is observed that the
equilibrium of the game changes whenever a competing peer leaves the
network. The bandwidth allocation for the existing peers increase
due to the leave of peers. This is in accordance with our analysis
given in Section \ref{Subsec-peersleave}. It is also observed that
the bandwidth assignment is proportional to the contribution value
of each peer at each equilibrium. This indicates that the proposed
incentive mechanism is robust to the dynamics of the P2P network,
and can always provide differentiated service to peers with
different contribution values. It is also interesting to observe
that the equilibriums of Example 5 are exactly the same as those of
the Example 4 for the same number of peers.  This is due to fact
that the uploader's utility obtained by selling the remaining
bandwidth to the remaining peers is larger than that obtained by
maintaining the current status in this example.

\section{Discussions and Future Work}\label{Sec-futurework}
\subsection{Competition among Multiple Uploaders}
In this paper, we consider a simple model where there is one
uploader and multiple downloaders. In reality, there may exist
multiple uploaders having overlapping data chunks. This implies that
there may exist a competition among these uploaders, which may
affect the pricing strategies of the uploaders. Our incentive
mechanism can be applied to this scenario with few modifications.

The presence of multiple uploaders induces a subgame that involves
the peers choosing the uploaders. This adds an additional "Step $0$:
Peers choose their uploader." to the proposed algorithms. Given any
set of choices by the peers, a Stackelberg game is induced at each
uploader, which can be solved for a unique SE according to the
analysis done in previous sections. Thus, the key issue is how to
choose the uploaders. When a new peer joins the network, it has to
choose one uploader from multiple uploaders that have its desired
data chunks. The prices at these uploaders observed by the newcomer
at this moment are fixed. Thus, it is reasonable for the newcomer to
choose the uploader with the lowest price.

However, it is worth pointing out that this scheme is in general
suboptimal. This is due to the fact that the price at the new SE
with the newcomer may not be the same as the price at the old SE
without the newcomer. Thus, it is possible that the newcomer
unilaterally deviate in its choice of the uploader to achieve a
higher utility. If we take this into consideration, the game will
become very complex and highly difficult to analyze. Thus, we would
like to delegate this to our future work.

\subsection{Trust Issues}
In this paper, we focus on designing an incentive mechanism for P2P
networks. However, it is worth pointing out that trust issues are
also very important for P2P systems. For example, the proposed
algorithms need the downloading peers to report their $c_i$'s and
$d_i$'s to the uploader. Malicious peers may misreport their credits
$c_i$'s and their types $d_i$'s to gain advantages against other
peers. For example, a malicious peer may deliberately reports a
bandwidth $\tilde{d}_i$ smaller than its real demand bandwidth $d_i$
to increase its priority ($c_i/d_i$) in obtaining bandwidth.

Another security issue is that malicious peers may deliberately
upload polluted data chunks to other peers. Without effective
measures to identify malicious peers, the polluted data chunks could
be disseminated to the whole network more quickly in a P2P network
with incentive mechanisms than that without incentive mechanisms.
This is due to the fact that peers are motivated to upload data
chunks to each other to earn points or monetary rewards in a P2P
system with incentive mechanisms. Without the ability to identify
malicious peers, peers are more likely to forward polluted data
chunks, consequently degrading the performance of the system.

To deal with these trust issues, trust management schemes are needed
to identify and defend against malicious peers. In other words,
incentive mechanisms must be used in trusted environments or
together with reliable trust management mechanisms. Though trust
management for P2P networks has been extensively studied in
literature
\cite{SKamvar_200305,YWang_200309,ASingh_200309,LXiong_200407,YLSun_200602,Rzhou_200704,kang2012fighting},
joint design of trust management and incentive mechanisms for P2P
networks remains unstudied. Due to the complexity and the lack of
space, we leave this as our future work.

\section{Conclusion}\label{conclusions}
In this paper, a credit-based incentive mechanism to stimulate the
cooperation between peers in a P2P streaming network is proposed.
Taking the peers' heterogeneity and selfish nature into
consideration, a Stackelberg game is designed to provide incentives
and service differentiation for peers with different credits and
connection types. The optimal pricing and purchasing strategies,
which can jointly maximize the uploader's and the downloaders'
utility functions, are derived by solving the Stackelberg game. The
Stackelberg equilibrium is shown to be unique and Pareto-optimal.
Then, two fully distributed implementation schemes are proposed and
studied. It is shown that each of these schemes has its own
advantages. The impact of peer churn on the proposed incentive
mechanism is then analyzed. It is shown that the proposed mechanism
can adapt to dynamic events such as peers joining or leaving the
network. Finally, several numerical examples are presented,
which show that the proposed incentive mechanism is effective in encouraging peers to cooperate with each other. 

\section*{ACKNOWLEDGMENT}
We would like to express our sincere thanks and appreciation to the
associate editor and the anonymous reviewers for their valuable
comments and helpful suggestions. This has resulted in a
significantly improved manuscript.

\end{document}